\documentclass[preprint,12pt]{elsarticle}

\usepackage{amssymb}
\usepackage{amsmath}

\usepackage{amsthm}
\usepackage{bm}

\begin{document}

\begin{frontmatter}



\title{Public and private beliefs under disinformation in social networks}


\author[1]{Diana Riazi}
\affiliation[1]{organization={Department of Computer Science, University College London},
            addressline={66-72 Gower Street}, 
            city={London},
            postcode={WC1E 6EA}, 
            country={UK}}

\author[2]{Giacomo Livan}
\affiliation[2]{organization={Department of Computer Science, University College London},
            addressline={66-72 Gower Street}, 
            city={London},
            postcode={WC1E 6EA}, 
            country={UK}}

\begin{abstract}
We develop a model of opinion dynamics where agents in a social network seek to learn a ground truth among a set of competing hypotheses. Agents in the network form private beliefs about such hypotheses by aggregating their neighbors' publicly stated beliefs, in an iterative fashion. This process allows us to keep track of scenarios where private and public beliefs align, leading to population-wide consensus on the ground truth, as well as scenarios where the two sets of beliefs fail to converge. The latter scenario --- which is reminiscent of the phenomenon of cognitive dissonance --- is induced by injecting `conspirators' in the network, i.e.,  agents who actively spread disinformation by not communicating accurately their private beliefs. We show that the agents' cognitive dissonance  non-trivially reaches its peak when conspirators are a relatively small minority of the population, and that such an effect can be mitigated --- although not erased --- by the presence of `debunker' agents in the network.
\end{abstract}



\begin{keyword}



\end{keyword}

\end{frontmatter}


\section{\label{sec:level1}Introduction}
We currently live in a paradoxical age of information. We have access to unprecedented amounts of information, yet our societies often fail to reach consensus on demonstrable facts \cite{mccright2011politicization}. This phenomenon has been widely studied in the opinion dynamics literature, leading to the identification of a number of potential factors leading to such an apparent paradox. One of such factors is the overwhelming amount --- both in terms of volume and diversity --- of information to which individuals are exposed to in online social networks (OSNs) \cite{rodriguez2014quantifying}. Even more importantly, the spread of disinformation and `fake news' has emerged as one of the most prevalent problems of today and remains an outstanding issue in the world of cyber and online security \cite{caramancion2020exploration}. In particular, studies have shown that more than two-thirds of people retrieve their news from social media platforms \cite{levy2017reuters}, thus the manner in which information spreads in OSNs is particularly pressing and has evidently emerged as a  threat to the stability of democracies and nations around the world \cite{farkas2019post,pantazi2021social}.  

A number of studies have illustrated the negative effects of the spread of disinformation on a population's \emph{collective} ability to reach consensus on a ground truth \cite{vosoughi2018spread,acemoglu2010spread}. In this paper, we seek to extend this line of research to the level of \emph{individual} agents in a population. Namely, we seek to illustrate the effects of disinformation on the formation of an agent's \emph{private} beliefs, and how these are converted into \emph{public} beliefs that are shared with others through a social network of interactions.

With this approach, we seek to understand the `microscopic' determinants of disagreement and divergence in social networks. In order to do that, we leverage a class of models known as Distributed Hypothesis Testing (DHT)~\cite{olfati2006belief}, and its connection to opinion dynamics and information diffusion. Originally developed in the context of sensor networks, DHT models typically describe a population of agents --- connected by a network of interactions --- seeking to learn a ground truth among a set of competing hypotheses \cite{lalitha2018social, olfati2006belief}.

The DHT literature has focused on designing models that reach consensus, in particular promoting exponentially fast convergence with the aim of possessing some form of robustness, via gauging a set of needed conditions \cite{nedic2017fast,shahrampour2013exponentially}. Incorporating further complexities, Hare \emph{et al.}~\cite{hare2020non} consider a non-Bayesian protocol designed to take into account agents' uncertainty regarding the statistical models (i.e., likelihood functions) which govern the state of world. 

Lalitha \emph{et al}.~\cite{lalitha2018social} put forward a model at the intersection of DHT and opinion dynamics, introducing in a DHT setting a learning rule \`a la DeGroot~\cite{degroot1974reaching} based on the iterative aggregation of \emph{local} information gathered from neighbors in a social network. It is precisely within this framework that private/public beliefs are explicitly modeled, as the agents gather and aggregate their neighbors' publicly stated opinions, but then proceed to form private beliefs based on them.

Similarly to~\cite{lalitha2018social}, much of the literature in opinion dynamics could be said to extend upon the archetypal DeGroot model~\cite{degroot1974reaching}, i.e., introducing additional complexities in a setting based on the iterative averaging/aggregation of information, and studying under what conditions the consensus typically reached in standard DeGroot models may be distorted into, e.g., a permanently polarized state. For instance, averaging models seek to show if and how Condorcet's Jury Theorem or wisdom of the crowds may be induced~\cite{buechel2015opinion}, with particular invoked behavioral qualities within agents~\cite{yildiz2013binary}. Moreover, it has been shown that social influence undermines the wisdom of the crowd~\cite{lorenz2011social}. At the same time, Becker \emph{et al}.~\cite{becker2017network} show network conditions where social influence may improve the accuracy of group estimates, despite the reduction in the diversity of opinions. Other classes of models seek to incorporate cognitive biases in the agents' information update process, e.g., explicitly modelling confirmation bias by having agents place more weight on opinions that already conform to their existing beliefs \cite{sikder2020minimalistic,allahverdyan2014opinion,del2017modeling}. Bounded confidence models instead impose that agents only interact if their opinions are similar enough \cite{lorenz2007continuous,del2017modeling}.

In the next section, we introduce the general structure of DHT models and detail its connection to opinion dynamics and social learning.

\section{\label{sec:level3}Social Learning and Distributed Hypothesis Testing}

In DHT models, agents seek to learn a ground truth via private signals, modelled as random draws from a distribution. The agents always receive signals from the distribution corresponding to the ground truth, but are unaware of that. In fact, their learning process aims at identifying from which distribution they are receiving signals among a number of different distributions. Each distribution may be thought as representing a competing hypothesis or narrative. In the following, we set our notation and formalise the learning rule we will employ throughout the paper.

We consider $N$ agents connected by a network $W$ of interactions (such that $W_{ij} > 0$ when agents $i$ and $j$ are connected and $W_{ij} = 0$ otherwise). The weight $W_{ij}$ represents the amount of influence that agent $i$ accepts from agent $j$. Because of this, the matrix $W$ is assumed to be row-stochastic, i.e. $\sum_{j=1} ^{N} W_{ij}=1$. Let us also consider a set of $N \times M$ multivariate distributions $f_i(X; \theta_{ik})$ ($i = 1, \ldots, N$, $k = 1, \ldots, M$), where $\theta_{ik}$ denotes the set of parameters that define the $k$-th distribution associated with agent $i$. To simplify notation, in the following we will drop the subscript $i$, as it will be clear from context to which agent the set of parameters $\theta_{ik}$ refers to.

We adopt the convention that --- for each agent --- the $M$-th distribution is the one corresponding to the ground truth. Time is discrete and denoted as $t=1,...,T$. At time $t = 0$ each agent is initialized with a random vector of private beliefs $\boldsymbol{q}_i^{(0)} = \left (q_i^{(0)}(\theta_1), \ldots, q_i^{(0)}(\theta_M) \right )$, such that $q_i^{(0)}(\theta_k) \geq 0, \ \forall k$ and $\sum_{k=1}^M q_i^{(0)}(\theta_k) = 1$. At each time step, the following occurs:

\begin{itemize}
    \item Each agent $i$ ($i = 1, \ldots, N$) receives a signal 
    $X_i^{(t)}$ as a random draw from the distribution corresponding to the ground truth, i.e., $X_i^{(t)}\sim f_i (\cdot ; \theta_M )$.
    \item Each agent $i$ performs a local Bayesian update on their current vector of private beliefs $\boldsymbol{q}_i^{(t)}$ to form a public belief vector $\boldsymbol{b}_i^{(t)}$ with components 
    \begin{equation} \label{eq:pub_beliefs}
        b_{i}^{(t)}(\theta_k) = \frac{f_{i}(X^{(t)}_{i}; \theta_k) \ q_i^{(t-1)}(\theta_k)}{\sum_{\ell = 1}^M f_i(X^{(t)}_{i};\theta_\ell) \ q^{(t-1)}_{i}(\theta_\ell)}
    \end{equation}            
    \item Each agent $i$ shares their public belief vector $\boldsymbol{b}_i^{(t)}$ with all their neighbors in the network, and similarly receives public belief vectors from each of them.
    \item Each agent $i$ updates their private belief vector $\boldsymbol{q}_i^{(t)}$ by averaging the log-beliefs they received from neighbors, i.e.,
    \begin{equation} \label{eq:pvt_beliefs}
q^{(t)}_{i}(\theta_k)=\frac{\exp \left (\sum_{j=1}^N W_{ij} \log b^{(t)}_{j}(\theta_k) \right)}{\sum_{\ell = 1}^M \exp \left (\sum_{j=1}^N W_{ij}\log b^{(t)}_{j}(\theta_\ell) \right )} \ .
    \end{equation}
\end{itemize}

Note that both Eqs.~\eqref{eq:pub_beliefs} and~\eqref{eq:pvt_beliefs} ensure that the public and private belief vectors of each agent remain correctly normalized as probability vectors at each time step.

It can be shown (see~\cite{lalitha2018social}) that as long as \emph{global distinguishability} is guaranteed, then the ground truth can be collectively learnt by all agents exponentially fast. Global distinguishability refers to the fact that at least one agent in the network is able to distinguish between a pair of competing hypotheses, i.e., for all $k \neq \ell$, there exists at least one agent $i$ such that $D_\mathrm{KL}(f_i(\cdot,\theta_k) || f_i(\cdot,\theta_\ell)) > 0$, where $D_\mathrm{KL}(\cdot || \cdot)$ denotes the Kullback-Leibler divergence.

\begin{figure}
    \centering
    \includegraphics[scale=.2]{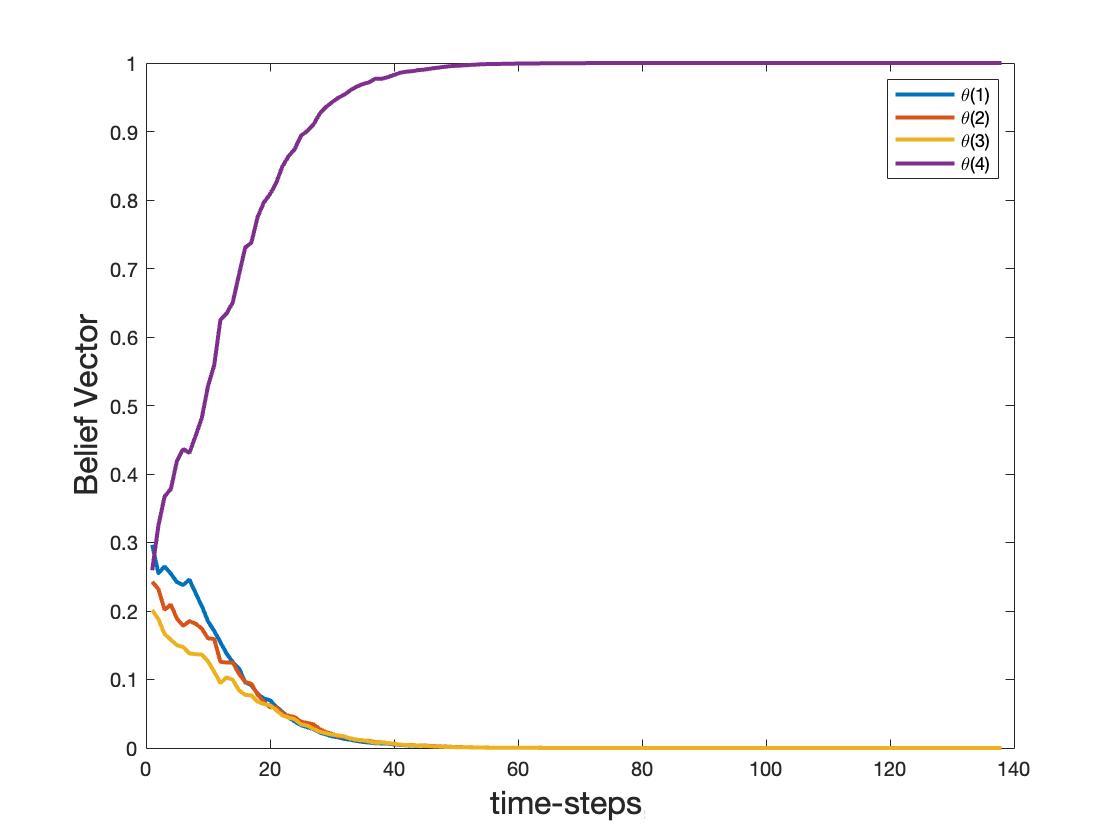}
    \caption{Temporal evolution of a randomly selected agent's private belief vector in a model with $N = 100$ agents and $M = 4$ hypotheses.}
     \label{fig:SL_simple}
\end{figure}

We illustrate the above result by simulating the learning rule on a fully-connected network with $N = 100$ agents and $M = 4$ hypotheses. The agents receive their signals from multivariate Gaussians, so that $\theta_k$ ($k = 1, \ldots, 4$) here represents the mean vectors and covariance matrices of each of such Gaussians. In order to ensure global distinguishability, we assume that each agent is only able to distinguish between pairs of competing hypotheses. For instance, we may have a certain agent $h$ with $f_h(\cdot; \theta_1) = f_h(\cdot; \theta_3)$ and $f_h(\cdot; \theta_2) = f_h(\cdot; \theta_4)$ (agent $h$ fails to distinguish between hypotheses $1$ and $3$, and between hypotheses $2$ and $4$), and  an agent $\ell$ with  $f_\ell(\cdot; \theta_1) = f_\ell(\cdot; \theta_2)$ and $f_\ell(\cdot; \theta_3) = f_\ell(\cdot; \theta_4)$ (agent $\ell$ fails to distinguish between hypotheses $1$ and $2$, and between hypotheses $3$ and $4$). In Fig.~\ref{fig:SL_simple} we show the temporal evolution of the private belief vector of a randomly chosen agent in the network. As it can be seen, the private belief in the ground truth quickly reaches one, whereas all other private beliefs go to zero. The same occurs for any other agent, leading to consensus on the ground truth.

\section{\label{sec:level4}Truthfulness and disinformation}
In the previous section we have introduced the behaviour of individual agents. Here, we begin to characterize the overall network's ability to learn the ground truth. To that end, we introduce the following quantity 

\begin{equation} \label{eq:truthfulnes}
    \tau(t) =\frac{1}{N}\sum_{i=1}^N q_i^{(t)}(\theta_M) \ ,
\end{equation}
which quantifies the average private belief placed by the agents on the ground truth (we recall that --- by convention --- we assume the $M$-th hypothesis to be the true one). We shall refer to the quantity in Eq.~\eqref{eq:truthfulnes} as \emph{truthfulness}. Thanks to the normalization of private belief vectors, truthfulness can also be expressed as the difference between one and the average private belief collectively placed on the $M-1$ wrong hypotheses:
\begin{equation*}
    \tau (t)= 1 - \frac{1}{N}\sum_{i=1}^N \sum_{\ell=1}^{M-1}
    q_i^{(t)}(\theta_\ell) \ .
\end{equation*}
When the aforementioned conditions on the global distinguishability of hypotheses are met, and the agents correctly learn the ground truth over time, one obviously has $\lim_{t \rightarrow \infty} \tau(t) = 1$, i.e., consensus on the ground truth.

Given that consensus on the ground truth is guaranteed under very mild conditions, our interest now shifts to what happens when those conditions are perturbed. First, we take a look at what happens after injecting \emph{conspirators}, i.e., agents that do not perform an update of their public beliefs as per Eq.~\eqref{eq:pub_beliefs} (and that do not update their private beliefs either). Specifically, conspirators maintain their public belief on one of the $M-1$ wrong hypotheses artificially high , therefore effectively spreading disinformation and affecting the `regular' agents' learning process. In particular, we impose that conspirator agents maintain a public belief vector $\boldsymbol{b}^{(t)} = (b_1,b_2,b_3,\ldots)$ with $b_1 \approx 1$ and $b_j \ll 1$ for $j = 2,\ldots, M$, that is, conspirator agents push one of the wrong hypotheses (which, without loss of generality, we assume to be the first one) at each time step.

\begin{figure}
    \centering
    \includegraphics[scale=.15]{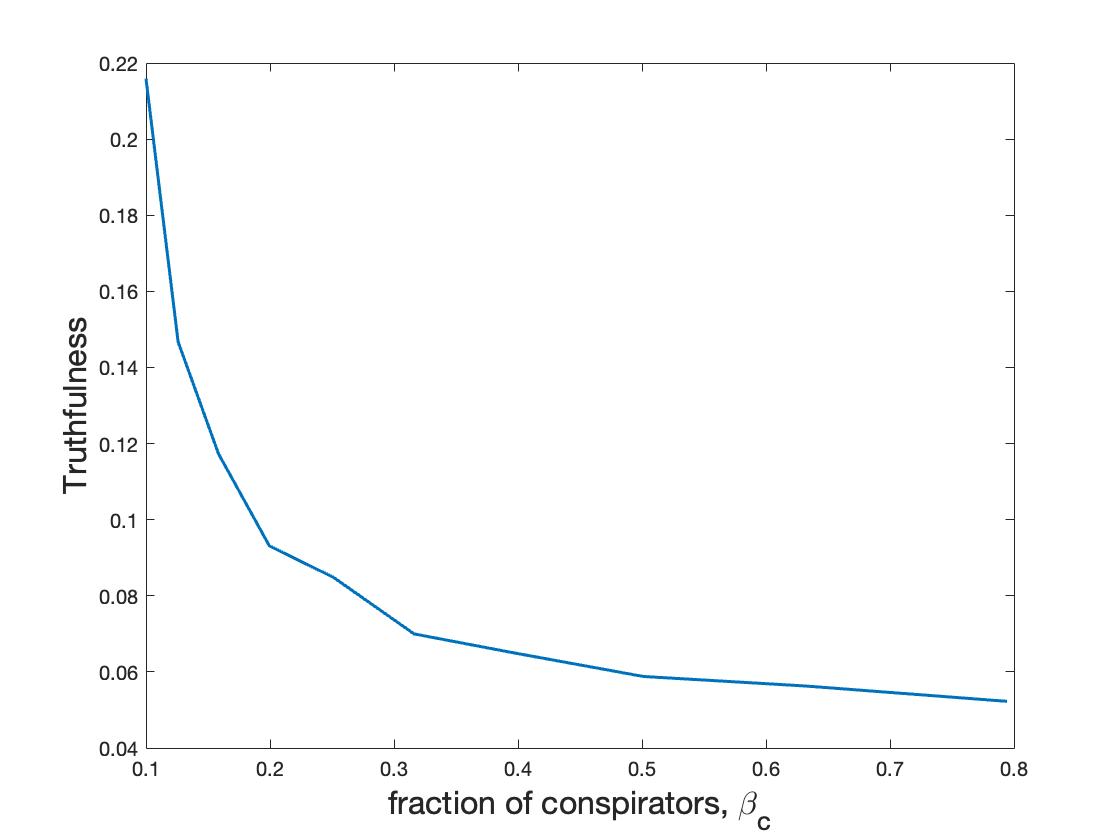}    
    \includegraphics[scale=.15]{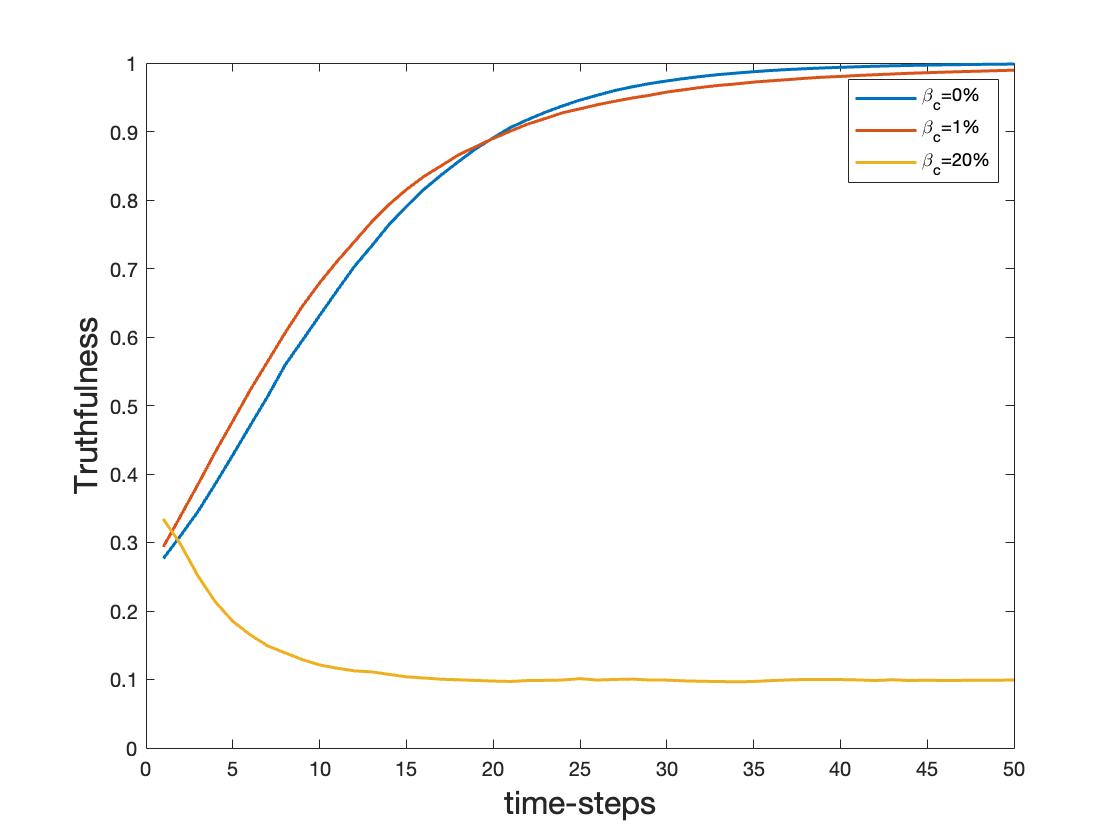}    
    \caption{Left panel: Average truthfulness in the regular agents sub-population as a function of the fraction of conspirators in the network. Right panel: temporal evolution of truthfulness with varying concentrations of conspirators in the network. }
    \label{fig:ft}
\end{figure}

Let us consider a fully connected network of $N=100$ agents with $M=4$ competing hypotheses. Fig.~\ref{fig:ft} (left panel) shows truthfulness in the long run --- where Eq.~\eqref{eq:truthfulnes} is only computed as an average over regular agents --- as a function of the concentration $\beta_{c}$ of (randomly placed) conspirators in the network. We observe an intuitive negative relationship. Namely, as the concentration of conspirators pushing a particular wrong hypothesis increases, the collective belief in the ground truth decreases and converges to nearly zero.

Additionally, in the right panel of Fig.~\ref{fig:ft}, we see the time evolution of the regular agents' belief in the truth, given some fixed concentration of conspirators. For a zero sub-population of conspirators, truthfulness ultimately converges to one, i.e., consensus on the ground truth. We also see that a perturbation of just a single conspirator ($\beta_{c} = 1/N$) imposed on a network population of $100$ agents is enough to deter a full consensus. Finally, given a larger proportion ($\beta_{c} = 0.2$) of conspirators, we see a noticeable reduction in truthfulness.

Given the result that truthfulness decreases monotonically with the proportion of conspirators, we may additionally consider whether every agent shares the same private belief in the true hypothesis or is there instead a non-trivial distribution of beliefs int he long run. In other words, does there exist a divergence among agents within networks induced by the presence of conspirators? 

\begin{figure}
    \centering
    \includegraphics[scale=.2]{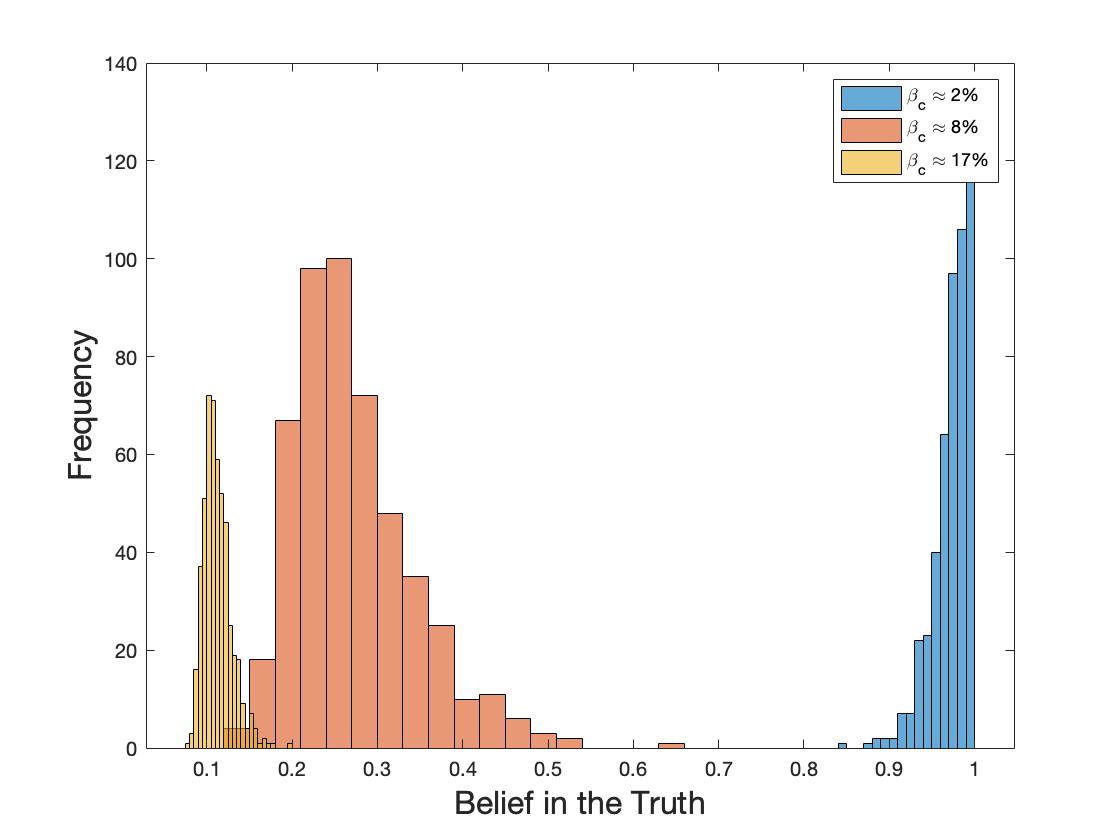}
    \caption{Steady state distributions of beliefs in the ground truth in the regular agents sub-population under varying concentrations of conspirators in the network ($N=300$, distributions obtained over $500$ distinct simulations)}
    \label{fig:dist_truth}
\end{figure}

As may be seen in Fig.~\ref{fig:dist_truth}, we observe the distributions of regular agents' private beliefs in the truth given some concentration of conspirators. Namely, with only $1/N$ conspirator present, we observe a rather skewed distribution, with most regular agents' private belief in the ground truth being equal or very close to one. Still, it is remarkable to see how the presence of a single conspirator is sufficient to render beliefs to significantly deviate from one for some agents. With $\beta_{c}=25/N$ we see a rather spread out distribution, with private beliefs in the ground truth ranging from $10\%$ to almost $80\%$, whereas with $\beta_{c}=50/N$ conspirators dominate and we see a significantly narrower distribution peaked around very low values.

\section{\label{sec:level5}A Model of Cognitive Dissonance}

In Psychology, cognitive dissonance (CD) refers to the mental toll experienced by an individual when faced with contradictory information~\cite{cooper2019cognitive}. In the context of our model, we assimilate CD to the difference (in absolute value) between what an agent privately believes and what they publicly express. Mathematically, the CD experienced by an agent $i$ at time $t$ regarding some hypothesis $\theta_k$ is
\begin{equation}
C^{(t)}_i(\theta_{k})= \left |q^{(t)}_i(\theta_k)-b^{(t)}_i(\theta_k) \right | \ .
\end{equation}

\begin{figure}
    \centering
    \includegraphics[scale=.2]{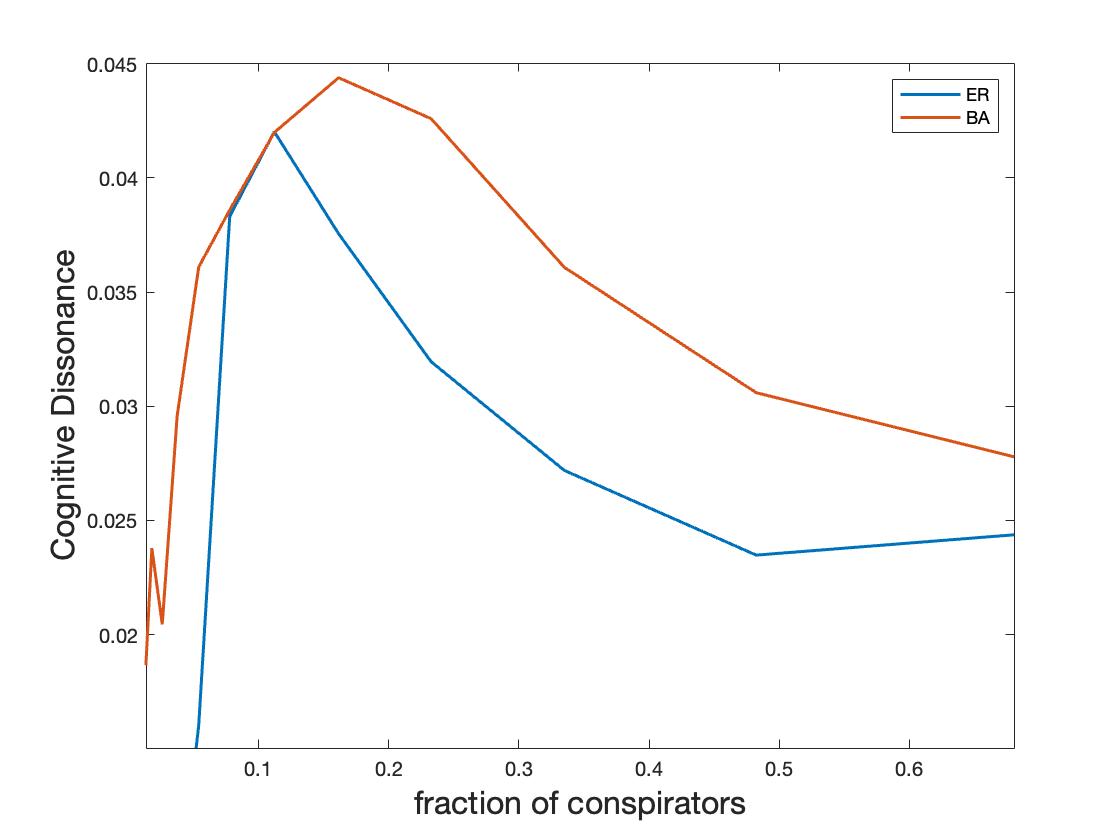}
    \caption{Cognitive dissonance as a function of the concentration of conspirators for a series BA and ER networks with average degree $k=100$ ($N=100$, averages obtained over $200$ simulations)}
    \label{fig:NM}
\end{figure}

We numerically simulate a series of Erdos-R\'enyi (ER) and Barabasi-Albert (BA) networks (with the same average degree) to consider the relationship which exists between the concentration of conspirators and the average CD experienced by regular agents, as well as to account for possible differences induced by network topology. In Fig.~\ref{fig:NM} we report the average CD experienced by regular agents with respect to the ground truth hypothesis ($C^{(t)}(\theta_{M})$) as a function of the proportion of conspirators in the network. The relationship is non-monotonic, and CD achieves its maximum in correspondence of a certain intermediate value $\beta_c \approx 10-15\%$. Notably, this result holds with little variation both in ER and BA networks.. This may seem counter-intuitive at first, as one may expect CD to be higher when regular agents are exposed to higher concentrations of conspirators. 

The above result becomes easier to interpret when considering the two extreme scenarios. When $\beta_c \rightarrow 0$ agents reach consensus on the ground truth, leading to both $q^{(t)}_i(\theta_M) \rightarrow 1$ and $b^{(t)}_i(\theta_M) \rightarrow 1$, and therefore $C^{(t)}_i(\theta_{M}) \rightarrow 0$ for all agents in the long run. Conversely, when $\beta_c \rightarrow 1$, the few regular agents present in the network are overwhelmed by the false information spread by conspirators, which similarly leads --- to some extent --- to the convergence of public and private beliefs, therefore suppressing CD. 

\begin{figure}
    \centering
    \includegraphics[scale=.2]{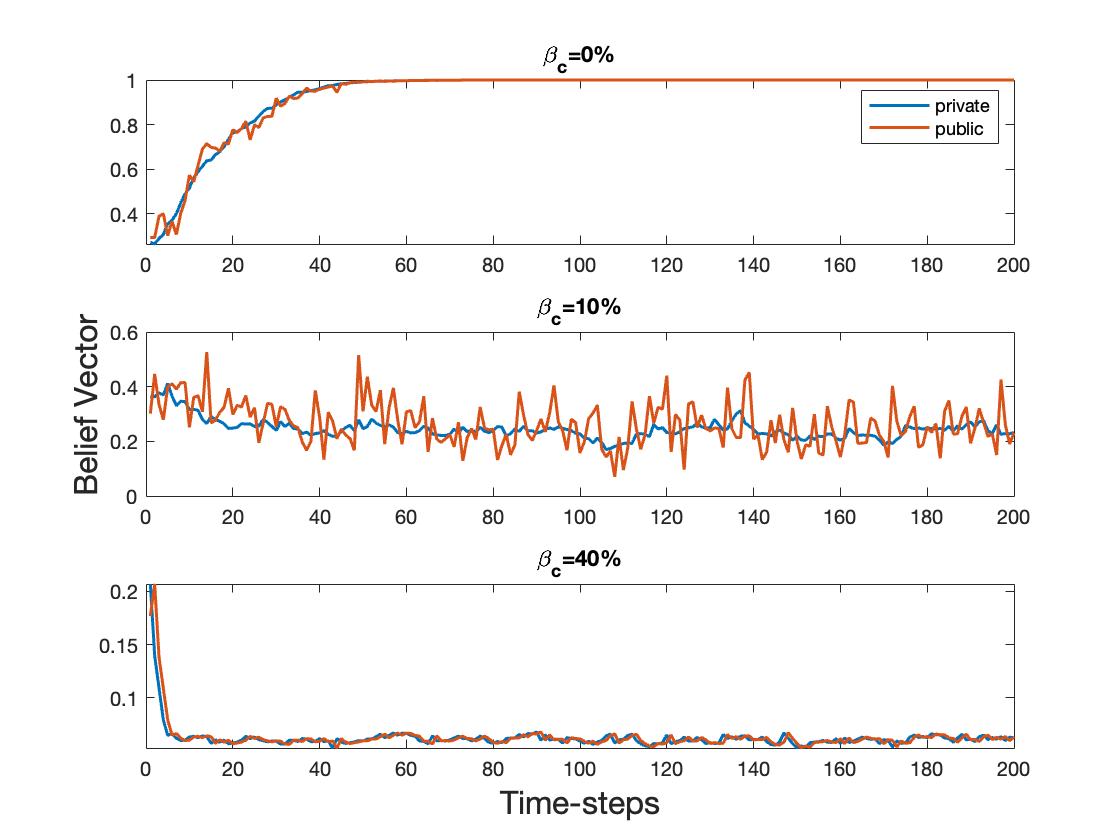}
    \caption{Temporal evolution of a randomly selected agent's public and private belief in the ground truth $\theta_M$ in a single simulation (with $N = 100$, $M = 4$). The concentrations of conspirators ($\beta_c$) are reported on the top of each plot.}
    \label{fig:public_private_beliefs}
\end{figure}

These results are further illustrated in Fig.~\ref{fig:public_private_beliefs}, where we report the temporal evolution of a randomly selected regular agent's public and private belief in the ground truth under different scenarios. We note that for low/high concentrations of conspirators, the agent's public and private beliefs quickly align. Conversely, for concentrations around the `critical' value $\beta_c \approx 10\% - 15\%$ we observe considerable oscillations of the agent's public beliefs, which do not die out in the long run. 

Given the consequences demonstrated regarding both truthfulness and CD, in the next section we explore whether such effects could be mitigated or suppressed by injecting `debunkers' in the network, i.e., agents who actively promote the ground truth.

\section{\label{sec:level6}Two-Tribe Model (the addition of `debunkers')}
We now consider a \textit{`two-tribe'} version of the model, akin to the dynamics of `conspirators' vs `debunkers', with the latter referring to an agent that actively pushes the ground truth $\theta_M$, where the parameter $\beta_d$ refers to the concentration of debunkers present within a network. We again model conspirators as agents with a static belief vector with $b_1^{(t)} \approx 1$ and $b_j^{(t)} \ll 1$ for $j = 2,\ldots,M$. In a similar fashion, we model debunkers to have a static belief vector with $b_M^{(t)} \approx 1$ and $b_j^{(t)} \ll 1$ for $j = 1,\ldots,M-1$.
As previously mentioned, we seek to understand whether the injection of debunkers in a social network may impact and potentially alleviate the effects of disinformation.

\begin{figure}
    \centering
    \includegraphics[scale=.2]{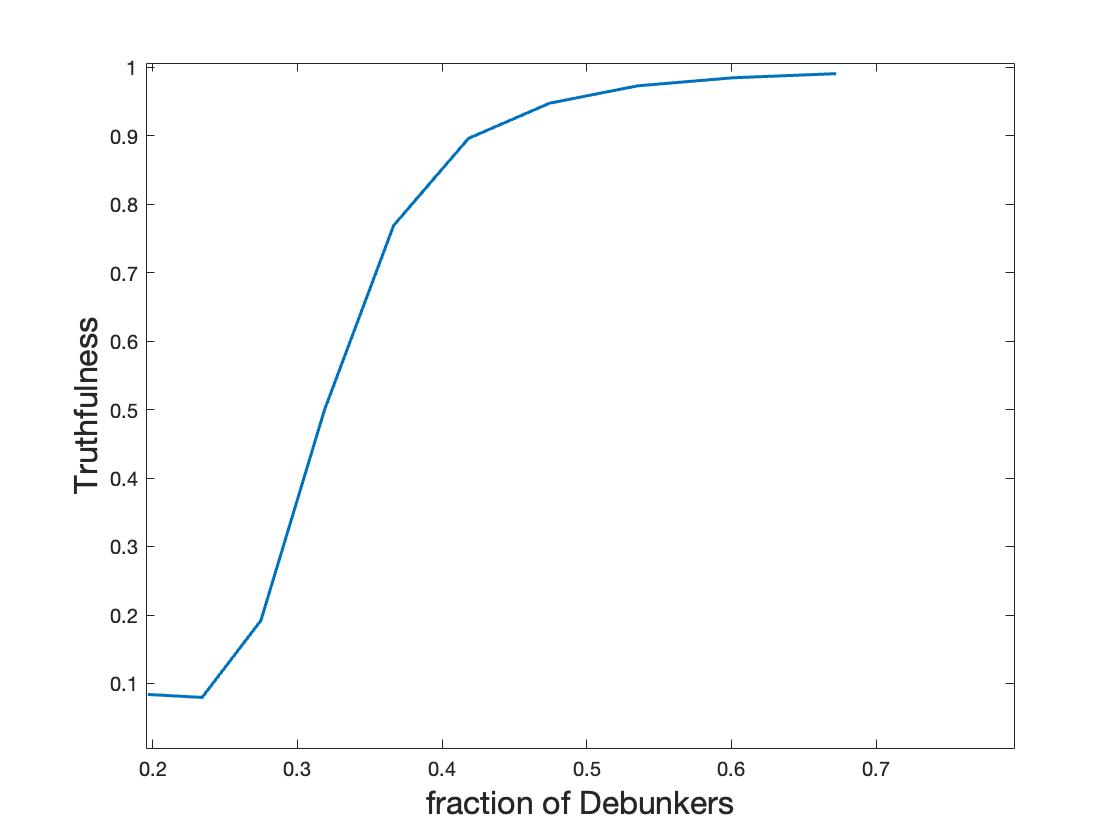}
    \caption{Average truthfulness in the regular agent sub-population as a function of the concentration of debunkers ($\beta_d$) in a model with $N=120$ agents and a fixed concentration of conspirators $\beta_c=25\%$.}
    \label{fig:TR3}
\end{figure}

We begin by considering the effects on truthfulness. In Fig.~\ref{fig:TR3}, it can be observed that, given a fixed fraction of conspirators, the collective belief in the ground truth approaches $1$ more closely as more debunkers are added to the network, as one would intuitively expect. In essence, this is the result which Fig.~\ref{fig:TR3} demonstrates, that is that truthfulness reverts to $1$, i.e. regular agents believe in the truth completely, as the concentration of debunkers increases. 

\begin{figure}
   \centering
   \includegraphics[scale=.15]{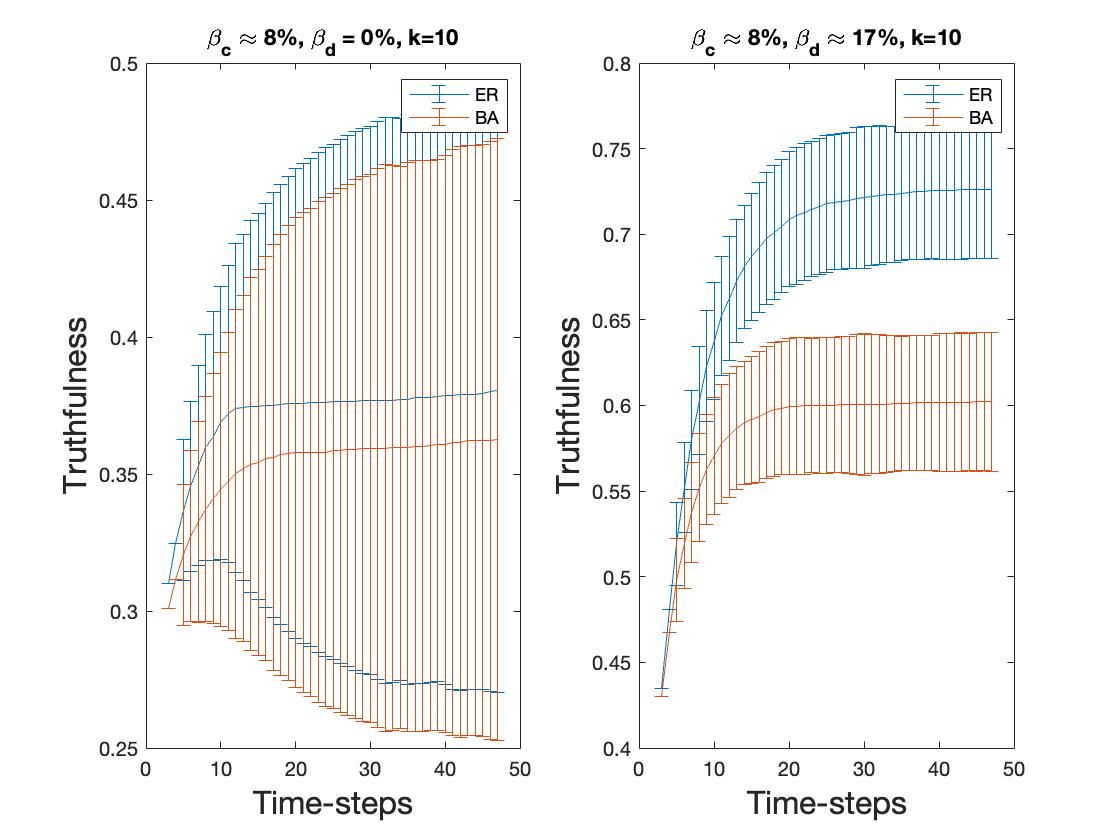}
   \includegraphics[scale=.15]{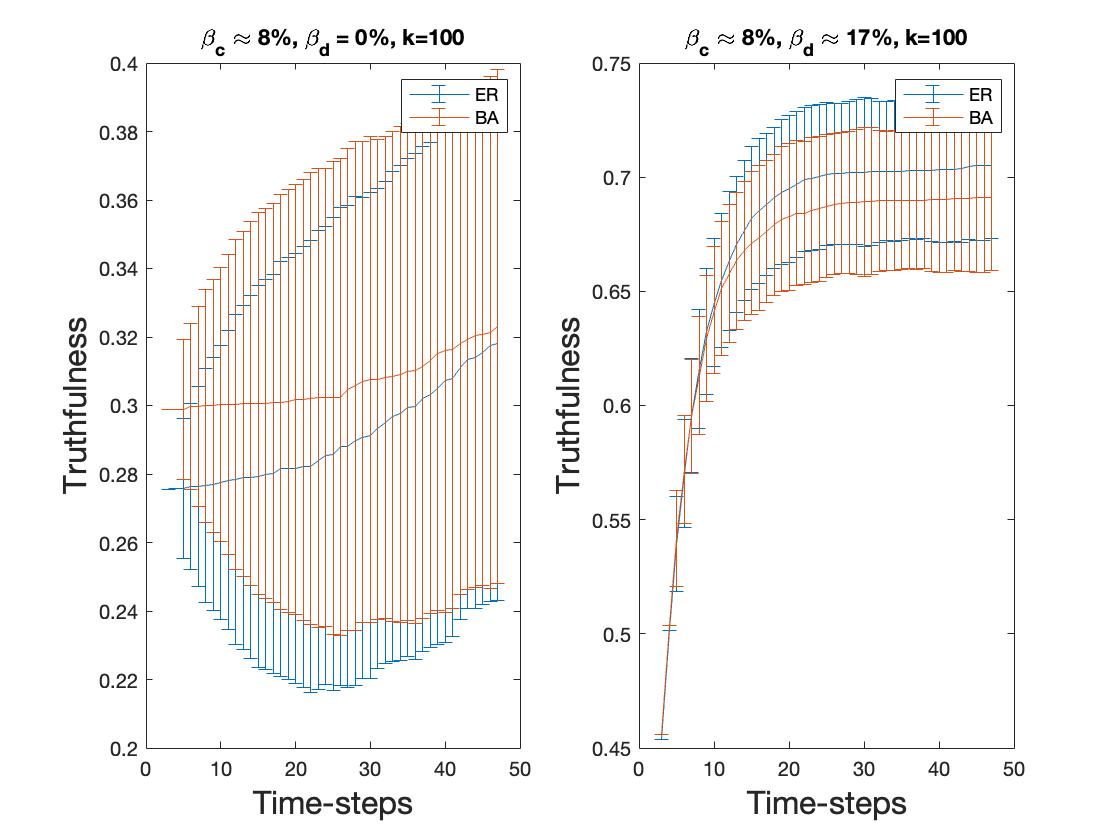}
    \caption{Comparison in truthfulness between ER and BA networks with average degrees equal to $k=10$ (left panels) and $k=100$ respectively (right panels), with a network size of $N = 300$. The concentrations of conspirators ($\beta_c$) and debunkers ($\beta_d$) are reported on top of each panel.}
    \label{fig:tr_10}
\end{figure}

\begin{figure}
  \centering
    \includegraphics[scale=.15]{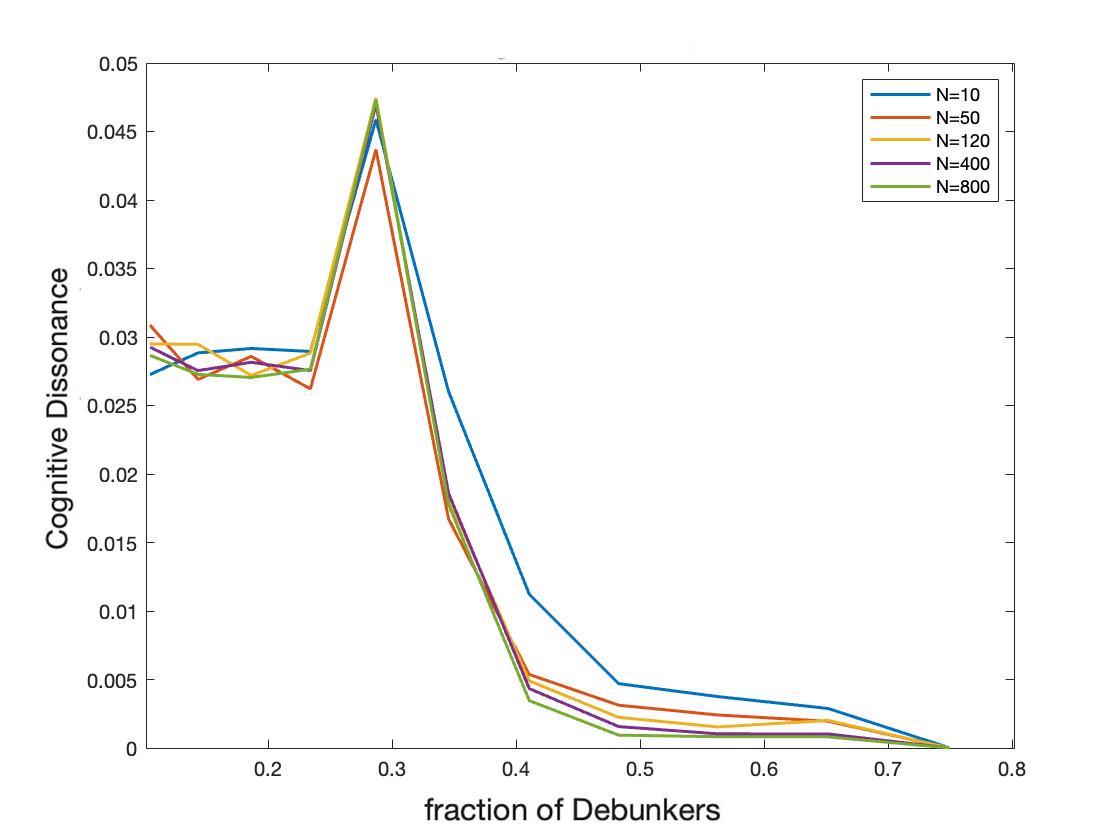}
    \includegraphics[scale=.15]{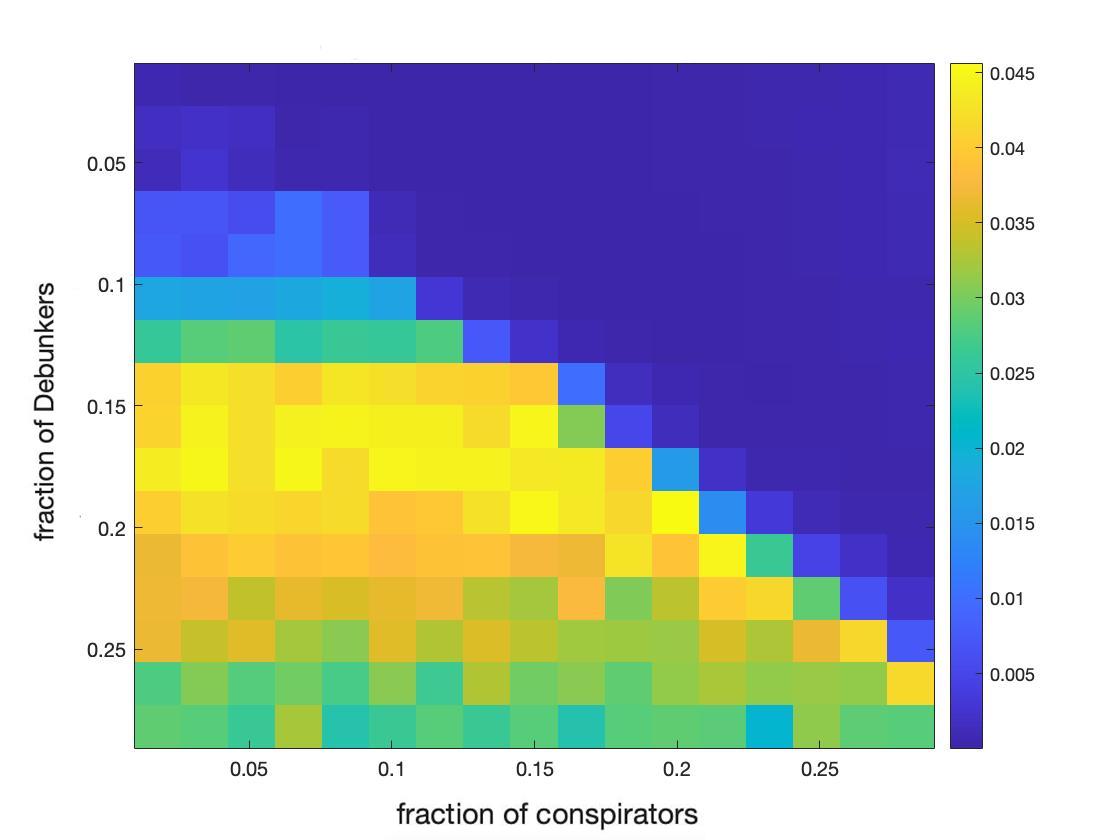}
    \caption{Left panel: Non-monotonicity in the behavior of CD as a function of the concentration of debunkers. Right panel: heatmap with steady state values of CD as a function of the concentrations of both conspirators and debunkers in the network.}
    \label{fig:CD}
\end{figure}

\begin{figure}
   \centering
    \includegraphics[scale=.2]{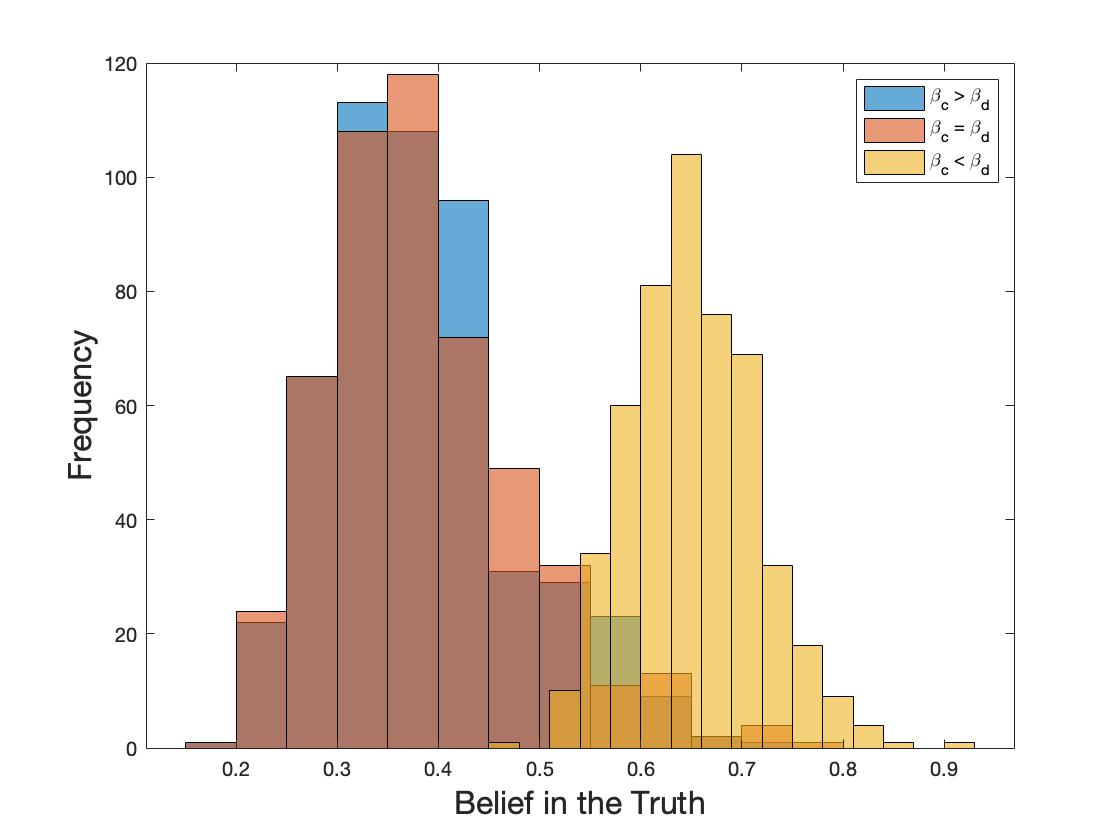}
    
    \caption{Distribution of private beliefs in the ground truth, with varying relative conspirator-debunker sub-populations, with a network population of $N=300$ for $500$ simulations}
    \label{fig:dist_deb}
\end{figure}

In Fig.~\ref{fig:tr_10}, we compare the temporal evolution of truthfulness on ER and BA networks, in order to determine what type of network is `better' at manufacturing truthfulness. We observe notable differences only for sparse networks, but as soon as the networks' average degree becomes large such differences cease to be statistically significant.

We now consider the effects on agents' cognitive dissonance induced by a sub-population of debunkers. As may be observed from Fig.~\ref{fig:CD} (left panel), CD does not decay monotonically upon injecting debunkers, but rather peaks when the sub-populations of conspirators and debunkers reach equal sizes. This result captures well the `backfire' effect on the downsides of debunking~\cite{redlawsk2002hot,nyhan2010corrections,zollo2017debunking}. In the right panel of Fig.~\ref{fig:CD} this result is presented more generally as a function of both the concentrations of conspirators and debunkers.

Lastly, in Fig.~\ref{fig:dist_deb} we report the steady state private belief distributions for the regular agents sub-population in scenarios where the concentration of conspirators exceeds the concentration of debunkers ($\beta_c > \beta_d$) and vice versa ($\beta_c < \beta_d$), as well as the scenario in which the two concentrations are equal ($\beta_c = \beta_d$). We see that only when debunkers are in larger numbers there is a sizeable shift to the right in the distribution, whereas the other two scenarios are fairly similar.

\section{\label{sec:level7}Conclusion}
In this paper, we put forward an opinion dynamics model aimed at describing the apparent tension between what individuals in a social network privately believe and what they publicly express. By relying on the DHT framework, we were able to produce both scenarios in which private and public beliefs quickly align for all individuals, as well as scenarios in which they remain permanently misaligned over time, which we relate to the psychological phenomenon of cognitive dissonance~\cite{cooper2019cognitive}. To the best of our knowledge, this aspect is novel in the opinion dynamics literature, although the distinction between public and private beliefs had minimalistically been attended to and implemented within models of social learning~\cite{shang2019resilient}.

Notably, the misalignment between private and public beliefs emerges only after introducing `conspirators' in a population of agents who process the information they receive from neighbors in a social network through a process of averaging \`a la DeGroot~\cite{degroot1974reaching}. We believe this result to be particularly interesting for two reasons. First, it provides a stylized model for the mental toll associated with the spread of disinformation in a social network. Second, this phenomenon does not respond monotonically to the proportion of conspirators in the population. Rather, it is maximized when conspirators are a relatively small minority of the population ($\approx 10\% - 15\%$). This is reminiscent of similar results on the disproportionate impact that small but dedicated minorities can have on the information aggregation capabilities of a society (see, e.g.,~\cite{livan2013leaders,sikder2020minimalistic,stella2018bots}).

Motivated by the above findings, we studied the model in a `two-tribe' setting, where we injected `debunker' agents with the aim of counteracting the effects induced by conspirators. Yet again, we found a non-trivial response in the model's behavior. We found the presence of debunkers to have a genuinely mitigating effect only when they begin to outnumber the conspirators. In fact, the CD experienced by regular agents in the network reaches its peak when the two sub-populations of conspirators and debunkers exactly match each other, a result which is very much reminiscent of the empirically demonstrated downsides of debunking attempts against the diffusion of disinformation and fake news~\cite{redlawsk2002hot,zollo2017debunking}.

In conclusion, it is worth mentioning that although we have continuously thought of the departure from the objective truth boasted by conspirator agents as `disinformation', i.e., possessing a malicious intent behind corrupting the truth, the conspirator-debunker dynamic has the potential for such a departure to cover cases of `misinformation' as well,  as conspirators may unknowingly or ignorantly be altering the truth.

\bibliographystyle{elsarticle-num} 
\bibliography{Riazi_Livan_bibliography}

\begin{thebibliography}{10}
\expandafter\ifx\csname url\endcsname\relax
  \def\url#1{\texttt{#1}}\fi
\expandafter\ifx\csname urlprefix\endcsname\relax\def\urlprefix{URL }\fi
\expandafter\ifx\csname href\endcsname\relax
  \def\href#1#2{#2} \def\path#1{#1}\fi

\bibitem{mccright2011politicization}
A.~M. McCright, R.~E. Dunlap, The politicization of climate change and
  polarization in the american public's views of global warming, 2001--2010,
  The Sociological Quarterly 52~(2) (2011) 155--194.

\bibitem{rodriguez2014quantifying}
M.~G. Rodriguez, K.~Gummadi, B.~Schoelkopf, Quantifying information overload in
  social media and its impact on social contagions, in: Proceedings of the
  international AAAI conference on web and social media, Vol.~8, 2014, pp.
  170--179.

\bibitem{caramancion2020exploration}
K.~M. Caramancion, An exploration of disinformation as a cybersecurity threat,
  in: 2020 3rd International Conference on Information and Computer
  Technologies (ICICT), IEEE, 2020, pp. 440--444.

\bibitem{levy2017reuters}
D.~Levy, N.~Newman, R.~Fletcher, A.~Kalogeropoulos, R.~K. Nielsen, Reuters
  institute digital news report 2014, Report of the Reuters Institute for the
  Study of Journalism (2017).

\bibitem{farkas2019post}
J.~Farkas, J.~Schou, Post-truth, fake news and democracy: Mapping the politics
  of falsehood, Routledge, 2019.

\bibitem{pantazi2021social}
M.~Pantazi, S.~Hale, O.~Klein, Social and cognitive aspects of the
  vulnerability to political misinformation, Political Psychology 42 (2021)
  267--304.

\bibitem{vosoughi2018spread}
S.~Vosoughi, D.~Roy, S.~Aral, The spread of true and false news online, science
  359~(6380) (2018) 1146--1151.

\bibitem{acemoglu2010spread}
D.~Acemoglu, A.~Ozdaglar, A.~ParandehGheibi, Spread of (mis) information in
  social networks, Games and Economic Behavior 70~(2) (2010) 194--227.

\bibitem{olfati2006belief}
R.~Olfati-Saber, E.~Franco, E.~Frazzoli, J.~S. Shamma, Belief consensus and
  distributed hypothesis testing in sensor networks, in: Networked Embedded
  Sensing and Control: Workshop NESC?05: University of Notre Dame, USA October
  2005 Proceedings, Springer, 2006, pp. 169--182.

\bibitem{lalitha2018social}
A.~Lalitha, T.~Javidi, A.~D. Sarwate, Social learning and distributed
  hypothesis testing, IEEE Transactions on Information Theory 64~(9) (2018)
  6161--6179.

\bibitem{nedic2017fast}
A.~Nedi{\'c}, A.~Olshevsky, C.~A. Uribe, Fast convergence rates for distributed
  non-bayesian learning, IEEE Transactions on Automatic Control 62~(11) (2017)
  5538--5553.

\bibitem{shahrampour2013exponentially}
S.~Shahrampour, A.~Jadbabaie, Exponentially fast parameter estimation in
  networks using distributed dual averaging, in: 52nd IEEE Conference on
  Decision and Control, IEEE, 2013, pp. 6196--6201.

\bibitem{hare2020non}
J.~Z. Hare, C.~A. Uribe, L.~Kaplan, A.~Jadbabaie, Non-bayesian social learning
  with uncertain models, IEEE Transactions on Signal Processing 68 (2020)
  4178--4193.

\bibitem{degroot1974reaching}
M.~H. DeGroot, Reaching a consensus, Journal of the American Statistical
  association 69~(345) (1974) 118--121.

\bibitem{buechel2015opinion}
B.~Buechel, T.~Hellmann, S.~Kl{\"o}{\ss}ner, Opinion dynamics and wisdom under
  conformity, Journal of Economic Dynamics and Control 52 (2015) 240--257.

\bibitem{yildiz2013binary}
E.~Yildiz, A.~Ozdaglar, D.~Acemoglu, A.~Saberi, A.~Scaglione, Binary opinion
  dynamics with stubborn agents, ACM Transactions on Economics and Computation
  (TEAC) 1~(4) (2013) 1--30.

\bibitem{lorenz2011social}
J.~Lorenz, H.~Rauhut, F.~Schweitzer, D.~Helbing, How social influence can
  undermine the wisdom of crowd effect, Proceedings of the national academy of
  sciences 108~(22) (2011) 9020--9025.

\bibitem{becker2017network}
J.~Becker, D.~Brackbill, D.~Centola, Network dynamics of social influence in
  the wisdom of crowds, Proceedings of the national academy of sciences
  114~(26) (2017) E5070--E5076.

\bibitem{sikder2020minimalistic}
O.~Sikder, R.~E. Smith, P.~Vivo, G.~Livan, A minimalistic model of bias,
  polarization and misinformation in social networks, Scientific reports 10~(1)
  (2020) 5493.

\bibitem{allahverdyan2014opinion}
A.~E. Allahverdyan, A.~Galstyan, Opinion dynamics with confirmation bias, PloS
  one 9~(7) (2014) e99557.

\bibitem{del2017modeling}
M.~Del~Vicario, A.~Scala, G.~Caldarelli, H.~E. Stanley, W.~Quattrociocchi,
  Modeling confirmation bias and polarization, Scientific reports 7~(1) (2017)
  40391.

\bibitem{lorenz2007continuous}
J.~Lorenz, Continuous opinion dynamics under bounded confidence: A survey,
  International Journal of Modern Physics C 18~(12) (2007) 1819--1838.

\bibitem{cooper2019cognitive}
J.~Cooper, Cognitive dissonance: Where we've been and where we're going,
  International Review of Social Psychology 32~(1) (2019) 7.

\bibitem{redlawsk2002hot}
D.~P. Redlawsk, Hot cognition or cool consideration? testing the effects of
  motivated reasoning on political decision making, Journal of Politics 64~(4)
  (2002) 1021--1044.

\bibitem{nyhan2010corrections}
B.~Nyhan, J.~Reifler, When corrections fail: The persistence of political
  misperceptions, Political Behavior 32~(2) (2010) 303--330.

\bibitem{zollo2017debunking}
F.~Zollo, A.~Bessi, M.~Del~Vicario, A.~Scala, G.~Caldarelli, L.~Shekhtman,
  S.~Havlin, W.~Quattrociocchi, Debunking in a world of tribes, PloS one 12~(7)
  (2017) e0181821.

\bibitem{shang2019resilient}
Y.~Shang, Resilient consensus for expressed and private opinions, IEEE
  Transactions on Cybernetics 51~(1) (2019) 318--331.

\bibitem{livan2013leaders}
G.~Livan, M.~Marsili, What do leaders know?, Entropy 15~(8) (2013) 3031--3044.

\bibitem{stella2018bots}
M.~Stella, E.~Ferrara, M.~De~Domenico, Bots increase exposure to negative and
  inflammatory content in online social systems, Proceedings of the National
  Academy of Sciences 115~(49) (2018) 12435--12440.

\end{thebibliography}

\end{document}